\documentclass[onecolumn,showpacs,superscriptaddress,aip,floatfix]{revtex4-1}
\usepackage{rotating}
\usepackage{amsmath}
\usepackage{color}
\usepackage{graphicx}
\usepackage{subcaption}
\usepackage{epsfig}
\usepackage{mathrsfs}
\usepackage{courier}
\usepackage{textcomp}
\usepackage[sort&compress]{natbib}

\newcommand{\kk}{{\bf k}}

\newcommand{\be}{\begin{equation}}
\newcommand{\ee}{\end{equation}}
\newcommand{\ben}{\begin{equation*}}
\newcommand{\een}{\end{equation*}}
\newcommand{\bea}{\begin{eqnarray}}
\newcommand{\eea}{\end{eqnarray}}
\newcommand{\bean}{\begin{eqnarray*}}
\newcommand{\eean}{\end{eqnarray*}}

\newcommand{\neel}{CNRS/Univ. Grenoble Alpes, Institut N\'eel, F-38042 Grenoble, France} 
\newcommand{\cinam}{Centre Interdisciplinaire des Nanosciences de Marseille, Aix-Marseille Universit\'{e},
Campus de Luminy, Marseille, 13288 Cedex 09, France} 
\newcommand{\torvergata}{University of Rome Tor Vergata, Rome, Italy}
\newcommand{\cea}{Laboratoire de Simulation Atomistique (L\_Sim), MEM, INAC, CEA, 38054 Grenoble Cedex 9, France}
\newcommand{\uga}{Universit\'{e} Grenoble Alpes, CS 40700, 38058 Grenoble Cedex, France}

\begin{document}
\title{Structural, electronic, and optical properties of the C-C complex in bulk silicon from first principles}
\author{Dilyara Timerkaeva}
\affiliation{\uga}
\affiliation{\cea}

\author{Claudio Attaccalite}
\affiliation{\neel}
\affiliation{\cinam}
\affiliation{\torvergata}
\author{Gilles Brenet}
\affiliation{\uga}
\affiliation{\cea}
\author{Damien Caliste}
\affiliation{\uga}
\affiliation{\cea}
\author{Pascal Pochet}
\affiliation{\uga}
\affiliation{\cea}

\begin{abstract}
The structure of the C$_i$C$_s$ complex  in silicon has long been the subject of debate. Numerous theoretical and experimental studies have attempted to shed light on the properties of these defects that are at the origin of the light emitting G-center.
These defects  are relevant for applications in lasing, and it would be advantageous to control their formation and concentration in bulk silicon. It is therefore essential to understand their structural and electronic properties.
In this paper, we present the structural, electronic, and optical properties of four possible configurations of the C$_i$C$_s$ complex in bulk silicon, namely the A-, B-, C-, and D-forms. The configurations were studied by density functional theory (DFT) and many-body perturbation theory (MBPT). Our results suggest that the C-form was misinterpreted as a B-form in some experiments. Our optical investigation also tends to exclude any contribution of A- and B-forms to light emission. Taken together, our results suggest that the C-form could play an important role in heavily carbon-doped silicon. 

\end{abstract}

\maketitle

\section{Introduction}
 Carbon, as an isovalent impurity to silicon, initially occupies a substitutional position (C$_s$). High-energy irradiation (electron, ion, proton, or gamma) creates fast-diffusing self-interstitials, some of which interact with carbon atoms and eject them from the  substitutional sites to create carbon interstitials ($C_i$). $C_i$ are mobile at temperatures above room temperature, and can thus interact with other impurities to form defect complexes. C$_i$C$_s$ is one of the defects induced by secondary irradiation. This complex is perhaps the most studied defect due to its rich physics and interesting structural, electronic, and optical features.
 
The pair is associated with a G-center which emits light at 0.97 eV (1280 nm).\cite{Berhanuddin2012a, Berhanuddin2012b} It was discovered in the 60's as a by-product in the silicon crystal caused by the radiation damage due to bombardment with high-energy electrons, ions and gamma rays. In recent years, significant efforts have been deployed to increase the concentration of G-centers \cite{Murata2011, Berhanuddin2012a, Berhanuddin2012b},  generally through surface alteration of silicon, followed by laser annealing. These technologies have promising applications in the development of a silicon laser.

\begin{figure*}
 \begin{center}
  \includegraphics[width=\textwidth]{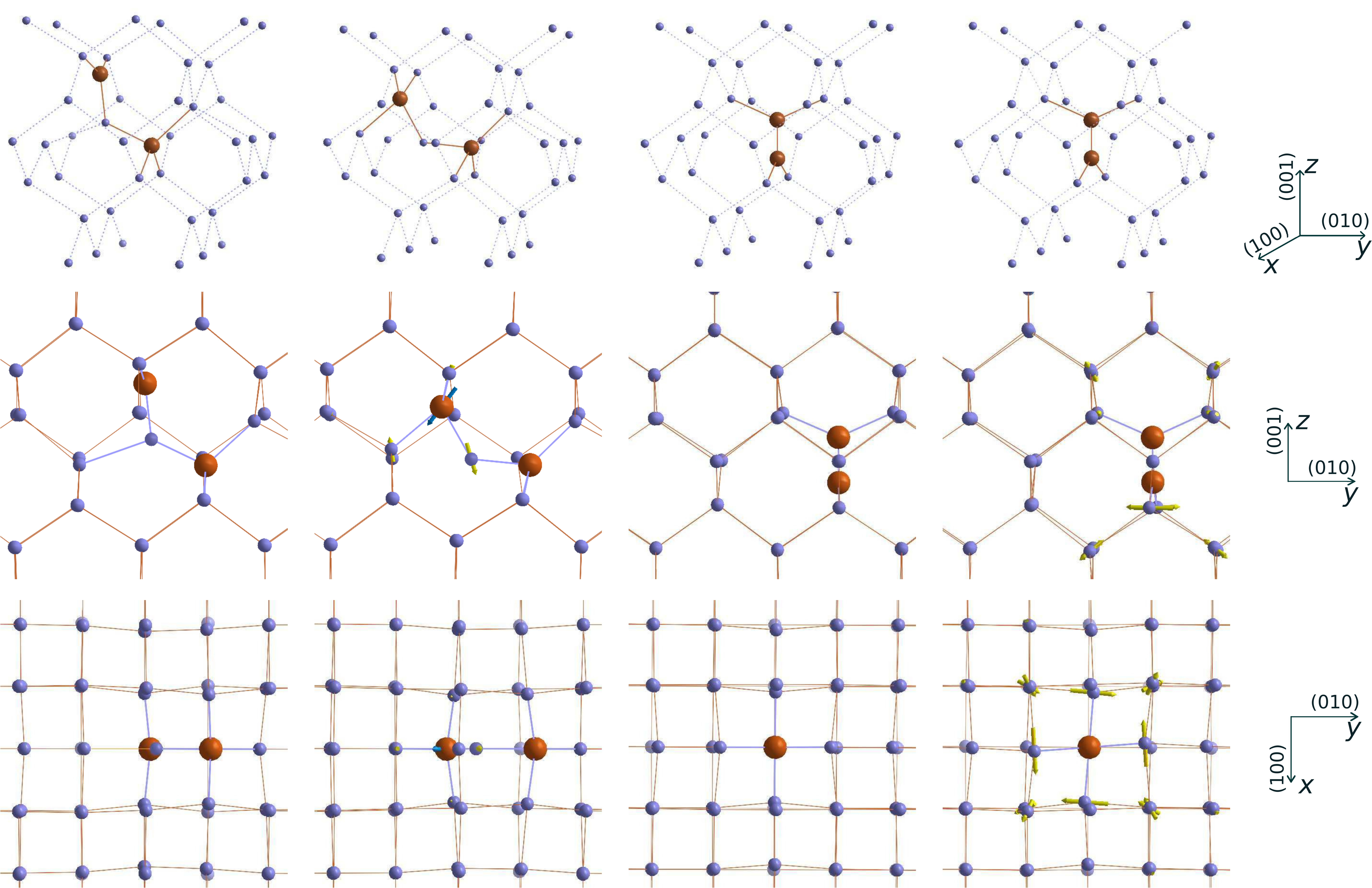}
 \end{center}
 \begin{flushleft}
  \begin{minipage}{0.22\textwidth}
   \subcaption{\label{fig:A-form} A-form}
  \end{minipage}
  \begin{minipage}{0.22\textwidth}
   \subcaption{\label{fig:B-form} B-form}
  \end{minipage}
  \begin{minipage}{0.22\textwidth}
   \subcaption{\label{fig:C-form} C-form}
  \end{minipage}
  \begin{minipage}{0.22\textwidth}
   \subcaption{\label{fig:D-form} D-form}
  \end{minipage}
 \end{flushleft}

 \caption{Schematic view (top line), "side" view (middle line), and "top" view (bottom line) of the four C$_i$C$_s$ complex forms. Blue spheres correspond to silicon atoms; red spheres correspond to carbon atoms. Yellow and blue arrows indicate distortion of the B-form with respect to the A-form and of the D-form with respect to the C-form. }
 \label{fig:forms}
 \end{figure*}
 
  The structure of the complex has been debated for a long time, and is the subject of numerous experimental and theoretical studies.\cite{Brower1974, ODonnell1983, Song1990, Lavrov1999, Leary1997, Capaz1998, Laiho2002, Wang2014, Wang2014jap}  
  Early Electron Paramagnetic Resonance (EPR) studies, conducted by Brower \textit{et al.},\cite{Brower1974} identified a signal, Si-G11, corresponding to a vacancy occupied by two carbon atoms in a positive charge state. The angular dependence of the Si-G11 Zeeman spectrum suggests that two carbon atoms lie in the \textlangle110\textrangle~plane, whereas the C-C bond is oriented along the (111) direction.  
%
%
  Subsequent experiments, based on optical detection of magnetic resonance (ODMR) studies of the 0.97~eV optical peak, linked the G-center to the C$_i$C$_s$ complex in its neutral state \cite{ODonnell1983}. These authors proposed an alternative model for the C$_i$C$_s$ complex, where the substitutional carbon atoms are separated by an interstitial silicon atom. Two modifications of the complex are possible: the A-form, where the interstitial silicon is in a three-bond configuration; and the B-form, where the Si is bound to two neighboring $C_s$ atoms.
  The bistable C$_i$C$_s$ complex (A- and B-forms in Figures~\ref{fig:A-form} and \ref{fig:B-form}) and its charged states have been experimentally studied.\cite{Song1990} Song \textit{et al.}\cite{Song1990} extensively analyzed bistable C$_i$C$_s$ complexes in p- and n-doped silicon using EPR, Deep Level Transient Spectroscopy (DLTS) and Photoluminescence spectroscopy (PL) techniques. Their results provided a complete configurational-coordinate energy diagram. The A-form was found to have lower energy for all the charge-states except neutral, for which the B-form had slightly lower energy. Later, the Localized Vibrational Modes (LVMs) of the bistable complex were determined by Infrared (IR) spectroscopy.\cite{Lavrov1999} 
  The spectra obtained for the B-form (540.4, 543.3, 579.8, 640.6, 730.4 and 842.4 $cm^{-1}$) and the A-form (594.6, 596.9, 722.4, 872.6, 953.0 $cm^{-1}$) agreed fairly well with the values determined by \textit{ab initio} calculations \cite{Leary1997, Capaz1998}. 
  
  In 2002, Laiho \textit{et al.} \cite{Laiho2002} used EPR to detect new low-symmetry configurations of a complex containing an interstitial silicon and two carbon atoms.  These signals, named Si-PT4 and Si-WL5, have yet to be characterized, however their presence proves the existence of additional forms of the C$_i$C$_s$ complex. The emergence of these new forms was linked to the cooling procedure employed in the experiments, and they were found to have varying magnetic properties.

  The precise geometry of carbon-pair configurations can be investigated by first principles studies. 
  A few attempts to theoretically determine their geometry have been made in recent years. By applying DFT calculations, Liu \textit{et al.} \cite{Liu2002} proposed a third configuration for the C$_i$C$_s$ complex (see Figure~\ref{fig:C-form}), the C-form, with two carbon atoms situated in a vacancy and oriented along the (100) direction. Although this configuration was found to be at least 0.2 eV more stable than the A and B forms for all charge states, it has never yet been experimentally observed.

  Most theoretical studies since the proposal of the C-form have investigated the three forms of C$_i$C$_s$ complexes \cite{Mattoni2002, Zirkelbach2011, Docaj2012} (see Figures~\ref{fig:A-form}, \ref{fig:B-form}, and \ref{fig:C-form}). However, their results were not always in agreement. 
  For example, Mattoni \textit{et al.} \cite{Mattoni2002} and Docaj \textit{et al.} \cite{Docaj2012} reported that binding of the A-form is as strong as that of the C-form, but that both forms are less stable than the B-form in a neutral state. In these two studies, non-spin-polarized calculations were performed. Zirkelbach \textit{et al.} \cite{Zirkelbach2011}, in contrast, calculated the stability of dicarbon pairs taking the spin into account. Their results identified the C-form as the most stable. In this article, we propose another stable configuration of the C$_i$C$_s$ complex, which we call the D-form. This form is produced by torsion of the C-form along the C-C bond axis (see Figure~\ref{fig:D-form}).

  Here, we present an extensive theoretical characterization of all four C$_i$C$_s$ complex forms from first-principles. The results will help to sort out which complex type is observed in different experimental growth procedures. We focused on both the ground state properties (geometries, binding energies) and the excited state properties (band gaps and optical absorption/emission spectra) of all four forms.
   
  The paper is organized as follows: in section \ref{methods} we summarize the computational methods used to calculate the atomic structure and the electronic and optical properties; in section \ref{comp_details} we present the actual numeric parameters used in the calculations; in section \ref{results} we present and discuss our results on the electronic, structural, and optical properties of the C$_i$C$_s$ complex forms, and we finally conclude in section \ref{conclusions}.

 \section{Methods}
 \label{methods}
  The C$_i$C$_s$ structures were embedded in 216-atom silicon supercells, applying periodic boundary conditions.
  Geometric optimizations were performed using BigDFT wavelet-based code\cite{Genovese2008} with Hartwigsen-Goedeker-Hutter pseudo-potentials\cite{Krack2005} and the GGA-PBE functional.\cite{PhysRevLett.77.3865} The locality of the wavelet basis-set and the appropriate parallelization of BigDFT ensure a very good efficiency for large supercells with localized orbitals, such as defect states.
 Structures were relaxed using the Fast Inertial Relaxation Engine (FIRE).\cite{Bitzek2006}
  To calculate the phonon modes, we employed a finite-difference method to the frozen phonon approximation. To accelerate computations, only the defect and its first-shell neighbors were precisely considered; remote atoms contributed to the dynamical matrix as unperturbed bulk atoms.

  The excited states were calculated by applying the many-body perturbation theory on a plane-wave basis set. Therefore, we
  took optimized structures from BigDFT and generated occupied and empty states\cite{noteopt} using the plane-wave based QuantumEspresso\cite{pwscf} code, norm-conserving Troullier-Martins pseudo-potentials\cite{troullier} and the GGA-PBE functional.\cite{PhysRevLett.77.3865}   \\
 The Kohn-Sham levels were then corrected by applying many-body perturbation theory to obtain the quasi-particle (QP) band structure and determine the system's optical response.
  The QP band structures were obtained within the $GW$ approach.\cite{aryasetiawan1998gw} Specifically, we used non-self-consistent $GW$ (denoted as $G_0W_0$) where the screened Coulomb potential, $W$, and Green's function, $G$, were built from the KS eigenstates  $\{\varepsilon_{n\kk}; |n\kk\rangle \}$ (where $\kk$ is the crystal wave vector and $n$ is the band index). The quasi-particle energies were then obtained from:
\be
\varepsilon^{\text{QP}}_{n\kk} = \varepsilon_{n\kk} + Z_{n\kk} \Delta \Sigma_{n\kk}(\varepsilon_{n\kk}).
\label{per}
\ee
In Eq.~\ref{per}
$$Z_{n\kk} = [1-\partial \Delta \Sigma_{n\kk}(\omega) / \partial \omega |_{\omega=\varepsilon_{n\kk}}]^{-1}$$
is  the re-normalization factor, and
$$\Delta \Sigma_{n\kk} \equiv \langle n\kk |\Delta \Sigma |n\kk \rangle,$$ 
where
$$\Delta \Sigma = \Sigma-V^{\text{xc}}$$ 
is the difference between  $\Sigma =GW$, the $GW$ self-energy, and $V^{\text{xc}}$, the exchange-correlation potential used in the KS calculation.\cite{aulbur1999quasiparticle} 

The optical-spectra were calculated by solving the Bethe-Salpeter equation (BSE):\cite{strinati}

\begin{align}
    (\varepsilon^{\text{QP}}_{c\kk}-\varepsilon^{\text{QP}}_{v\kk})A^{s}_{vc\kk}&+\nonumber \\ \sum_{v'c'\kk'}\langle vc\kk | K_{eh} | v'c'\kk' \rangle A^{s}_{v'c'\kk}&=
    \Omega^{s} A^{s}_{vc\kk}.
    \label{eq:bse}
\end{align}    
Here, electronic excitation was expressed in an electron-hole pair basis $|vc\kk\rangle$ corresponding to transitions at a given $\kk$ from a state in the valence band ($v$) with energy $\varepsilon^{\text{QP}}_{v\kk}$ (hole), to a conduction-band ($c$) state with energy $\varepsilon^{QP}_{c\kk}$ (electron). $A^{s}_{vc\kk}$ are the expansion coefficients of the excitons in the electron-hole basis, and the $\Omega^{s}$ are the excitation energies for the system.     
With spin-polarized defects, we performed spin-polarized calculations both at the GW and BSE levels. 

\begin{figure}[ht]
\centering
\includegraphics[width=0.9\textwidth]{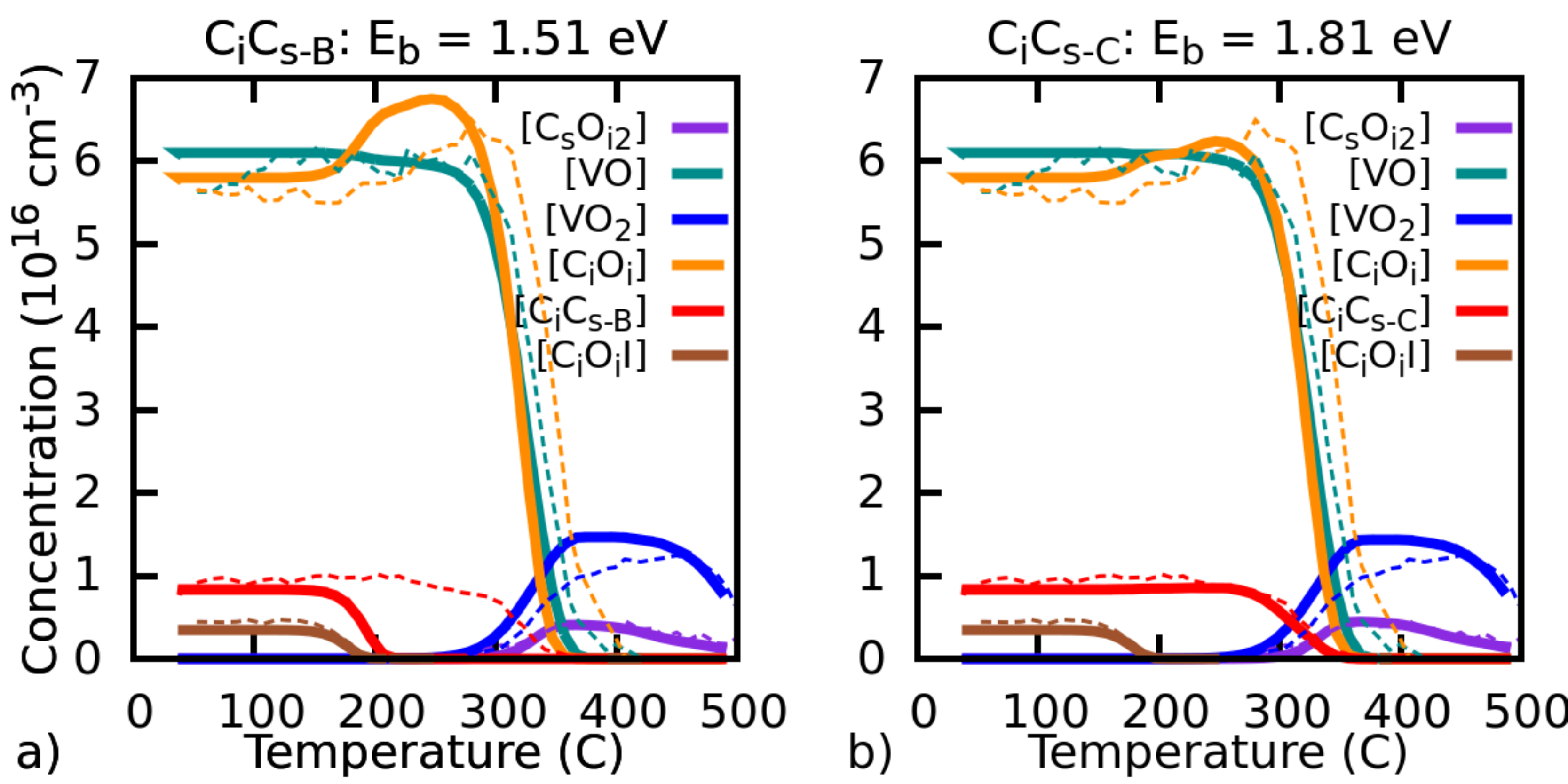}
\caption{[Color online] The C-form predominates in the experiments under consideration. Isochronal annealing simulations of carbon/oxygen related defects in silicons were compared to experimental observations\cite{Sgourou2013, Londos2013}. In experiments, carbon-containing silicon was first exposed to electron irradiation to provoke the appearance of Frenkel pairs. Fast-migrating vacancies and interstitials led to the formation of defect complexes. The concentrations of these formations were examined as a function of temperature. 
%
%
Continuous lines correspond to simulated curves; dashed lines represent experimental data. \textit{a)} KMAL simulations of the B-form of the C$_i$C$_s$ complex: simulation data fail to reproduce experimental data. \textit{b)} KMAL simulations of the C-form of the C$_i$C$_s$ complex. Experimental and simulated data show good matching. 
\label{fig:kmal}}
\end{figure}

\section{Calculation details}
 \label{comp_details}
In this section, we report the convergence parameters entering into the different calculations discussed in the previous section. 

As recently shown by Wang \textit{et al.} \cite{Wang2014, Wang2014jap}, the formation energy of the neutral states of A-, B-, and C-forms are the lowest among all charge states for a wide range of Fermi energy levels. Therefore, we focused on the neutral states of C$_i$C$_s$. Structures were optimized using a 0.42-Bohr grid spacing on the wavelet mesh, and a cutoff of 80~Ry in plane-wave to generate the KS wave-functions. In the  $G_0W_0$ calculation, 1500 bands were used to expand the Green's functions and calculate the screened interaction $W$, 30,000~$G$-vectors to expand the Kohn-Sham orbitals, 2-Ha cutoff for the $G-$vectors contributing to the dielectric constant. The dielectric contribution to the definition of $W$ was calculated using a double-grid technique to integrate the Brillouin Zone with a single k-point (gamma point) for the matrix elements, and a $2 \times 2 \times 2$ shifted grid for single-particle energies, see Kammerlander \textit{et al.}~\cite{kammerlander2012speeding} for details. The same double-grid technique was used to calculate the optical response. For the BSE, we used the static part of the screening calculated in the GW, 100 valence bands and 100 conduction bands to calculate the optical absorption, and 15,000 $G$-vectors to describe the Kohn-Sham wave-functions. Then, the BSE equation was also inverted and interpolated on the double grid.\cite{kammerlander2012speeding} All the GW and BSE calculations were performed in Yambo code.\cite{yambo}

\section{Results and discussion}
\label{results}
Here, we will address the structural, vibrational, electronic and optical properties of the different C$_i$C$_s$ complexes.


\begin{figure}[ht]
 \centering
 \begin{minipage}{8cm}
  \includegraphics[width=\textwidth]{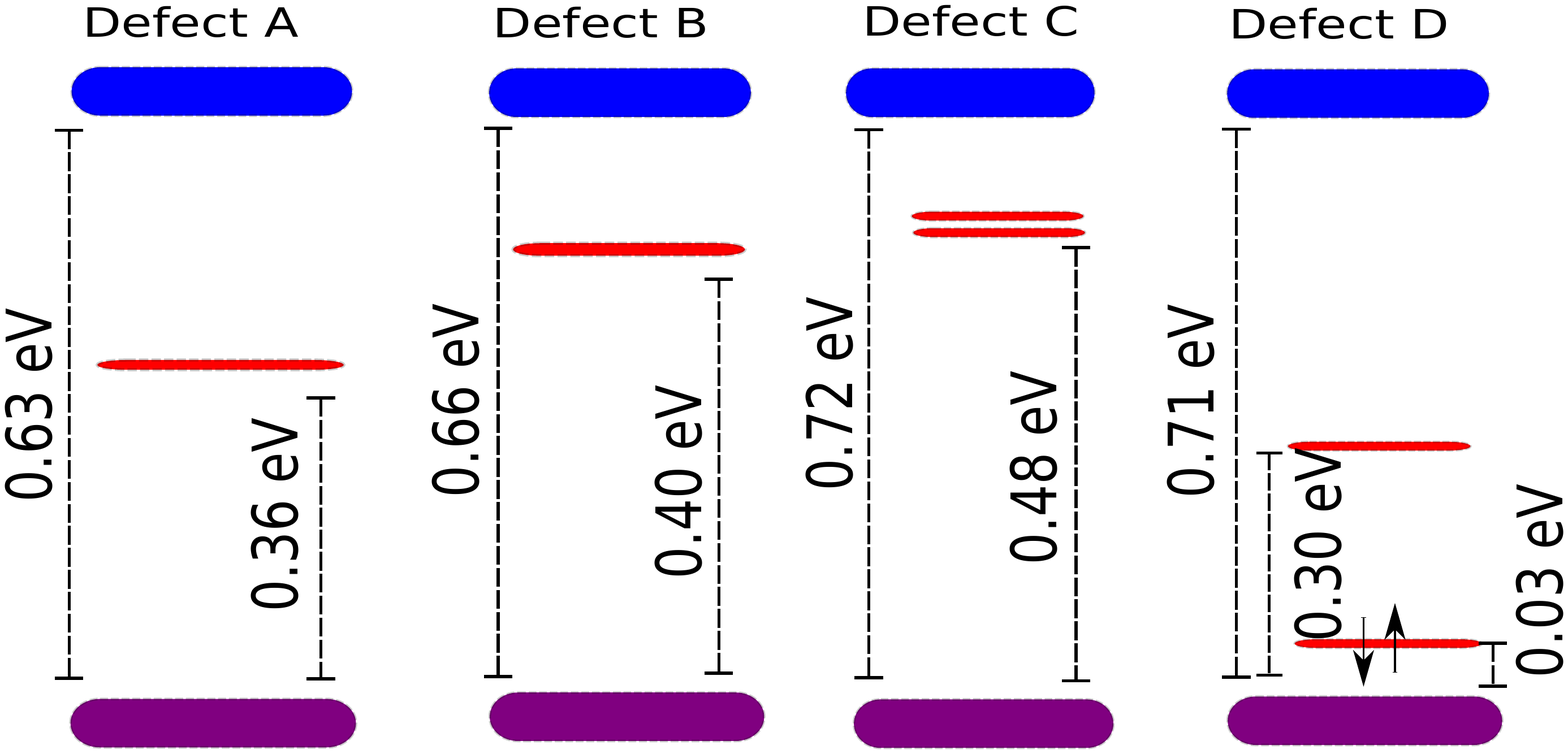}
 \subcaption{\label{fig:pbebands} DFT level}
 \end{minipage}\\
 \begin{minipage}{8cm}
  \includegraphics[width=\textwidth]{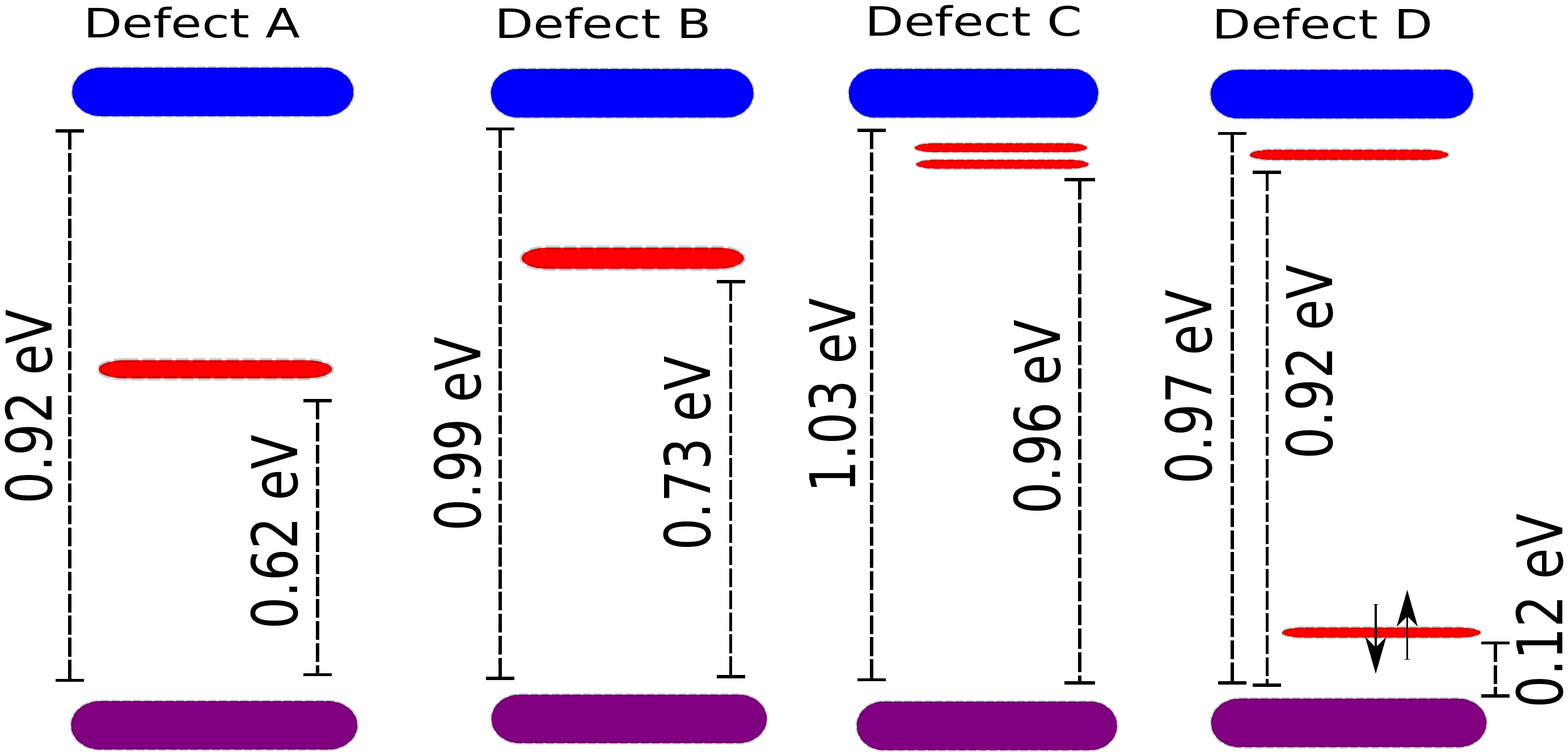}
 \subcaption{\label{fig:gwbands} G$_0$W$_0$ level} 
 \end{minipage}
 \caption{[Color online] Schematic band structure of the defects levels in the different complexes. Purple and blue colors correspond to valence and conduction bands, while red color corresponds to defect levels. All defects levels are empty except in the D case where there is a fully occupied level in the gap, indicated by the two arrows. }
\end{figure}


 \paragraph{Structures and binding energies of the various forms of the CiCs complex}
In Figures \ref{fig:A-form} - \ref{fig:D-form}, the four optimized configurations of the C$_i$C$_s$ complex are shown (A-, B-, C-, and D- forms). In the A-form, carbon atoms occupy neighboring lattice sites, while the Si atom is an interstitial bonded to both carbon atoms and one silicon. The B-form is similar to the A-form, but the interstitial Si only bridges the two carbon atoms and is not bonded to another Si. The third configuration, the C-form, consists of two carbons in a vacancy aligned in a (100) crystallographic direction. The D-form is a slight variation on the previous configuration: all Si atoms bonded to two C atoms are slightly twisted around the C-C axis compared to the C-form. The D-form was obtained through non-spin-polarized geometry optimization. The initial configuration for this optimization was chosen as slightly different from the C-form. Thus, we can conclude, that the D-form is another local minimum, which is similar to the C-form but with almost 0.4 eV higher energy. 

The twisted shape of the D-form could be the result of the rotation of the $\pi$-orbitals of two C atoms to form a $\pi$ bond (see Figures \ref{fig:C-form} and \ref{fig:D-form}). In the C-form, the two corresponding $\pi$-orbitals are perpendicular to each other; each is occupied by a single electron, making this complex paramagnetic.
  
All four forms of the C$_i$C$_s$ complex can be created from identical ingredients, \textit{i.e.} from mobile interstitial carbon and immobile carbon in a substitutional site. Depending on the topology of the reaction, either A and B or C and D forms can emerge. 
In a neutral state, A will directly transform to B, as B is more stable. The A to B transformation barrier has been estimated to be as low as ~0.1 eV \cite{Zirkelbach2011}. Transformation from either A (or B)  to C and back is less likely as the transformation barrier is estimated to be up to 2-3 eV \cite{Docaj2012}.  
The kinetics of C$_i$C$_s$ complex formation and reorientation, \textit{i.e.} various migration barriers, is beyond the scope of the present study. Interested readers can find information in Zirkelbach \textit{et al.}\cite{Zirkelbach2011}.

 The binding energies of the complex were calculated as 
 $E_b(C_iC_s) = - E_{tot}(215SiC_iC_s) - E_{tot}(216Si) + E_{tot}(215SiC_s) + E_{tot}(216SiC_i)$ and the values obtained are listed in Table \ref{tab:bindingenergy}. 
 The C-form was found to be the most stable. 
 These results contradict some recent theoretical studies \cite{Docaj2012, Mattoni2002}, however they concur with findings reported by Zirkelbach \textit{et al.} \cite{Zirkelbach2011} and Liu \textit{et al.} \cite{Liu2002}. We demonstrate the crucial role of the spin in complex respective stabilities, as it increases the binding energy by about 0.17 eV compared to non-spin polarized calculations (0.2 eV in Ref.\cite{Zirkelbach2011}). 
 The D configuration has a binding energy of 0.88 eV, which is close to that of the A-form. 

 \begin{table}[h]
 \centering
 \begin{tabular}{c|ccccccc}
 \hline 
 E$_b$, eV & {\footnotesize This work} & {\footnotesize Ref.\onlinecite{Zirkelbach2011}} & {\footnotesize Ref.\onlinecite{Docaj2012}} & {\footnotesize Ref.\onlinecite{Mattoni2002}} & {\footnotesize Ref.\onlinecite{Leary1997}} & {\footnotesize Ref.\onlinecite{Capaz1998}} &  {\footnotesize Ref.\onlinecite{Liu2002}} \\
  \hline 
  {\footnotesize A-form}       & 0.86 & 0.93 & 0.92 & E$_{b0}$     & E$_{b0}$-0.35 & E$_{b0}$-0.11 & E$_{b0}$-0.2 \\
  {\footnotesize B-form}       & 0.93 & 0.95 & 1.28 & E$_{b0}$-0.4 & E$_{b0}$      & E$_{b0}$      & E$_{b0}$-0.2 \\
  {\footnotesize C-form} & 1.11 &      & 0.90 & E$_{b0}$-0.2 &        &        &       \\
  {\footnotesize C-form(SP)}  & 1.28 & 1.28 &      &       &        &        & E$_{b0}$     \\
  {\footnotesize D-form} & 0.88 &      &      &       &        &        &       \\
  \hline
 \end{tabular}
 \caption{Binding energies in eV for four configurations of C$_i$C$_s$. References \cite{Mattoni2002}, \cite{Leary1997}, \cite{Capaz1998}, and \cite{Liu2002} reported relative values of their binding energies. E$_{b0}$ indicates the binding energy of the most stable configuration within each study. SP indicates spin-polarized calculations.}
 \label{tab:bindingenergy}
\end{table}

 Generally speaking, all these forms may be present in a heavily carbon-doped silicon, and their relative concentrations should depend on their binding energies if thermal equilibrium is reached, or otherwise on the thermal history of the sample. It should also be noted that formation kinetics can significantly affect the balance between the four complex concentrations.  That is why, even if the A- and B-forms are indeed less stable than the C-form, they may nevertheless be present in the sample and could be detected by various experimental techniques, such as IR and EPR spectroscopy. 

To reproduce the sample preparation process, we simulated the experimental sample preparation, tracking concentrations of various compound point defects in silicon exposed to electron irradiation. The concentrations of defects were extracted from Kinetic Mass Action Law (KMAL) simulations\cite{Brenet2015}. KMAL is a theoretical model based on a rate theory, which we use to reproduce temperature-driven diffusion-limited reactions. To account for the temperature effect, activation energies derived from \textit{ab initio} computation were used\cite{Sgourou2013}. The lattice type, the geometry of each defect, and their binding energies are the main factors affecting diffusion, defect formation and subsequent stability.

Fig. \ref{fig:kmal} shows the concentrations of various carbon-related defects in silicon produced during isochronal annealing (20 min) up to 500 \textcelsius. KMAL simulation results were compared to experimental data\cite{Sgourou2013, Londos2013}. We performed two simulations varying a single parameter: the C$_i$C$_s$ binding energy. Fig. \ref{fig:kmal}a shows results for the B-form of C$_i$C$_s$, whereas Fig. \ref{fig:kmal}b shows results for the C-form of C$_i$C$_s$.  
 
 The two simulations highlight the main features of the experiment: species, their concentrations, and the reaction temperatures. 
 When considering the C-form, the simulation reproduces the following two experimental parameters:
 the C$_i$C$_s$ dissociation temperature of 280-300 \textcelsius; 
 and the experimentally-observed evolution of the C$_i$O$_i$ pair. This evidence favors the existence of the C-form.
 With the B-form, decay of C$_i$C$_s$ pairs at lower temperatures (190 \textcelsius) releases mobile C$_i$ species, causing the C$_i$O$_i$ concentration to increase just before its dissociation. This increase is not seen experimentally, adding further proof that the C-form is the one detected in this experiment.

\begin{figure}[ht]
 \centering
 \begin{minipage}{0.8\textwidth}
  \includegraphics[width=\textwidth]{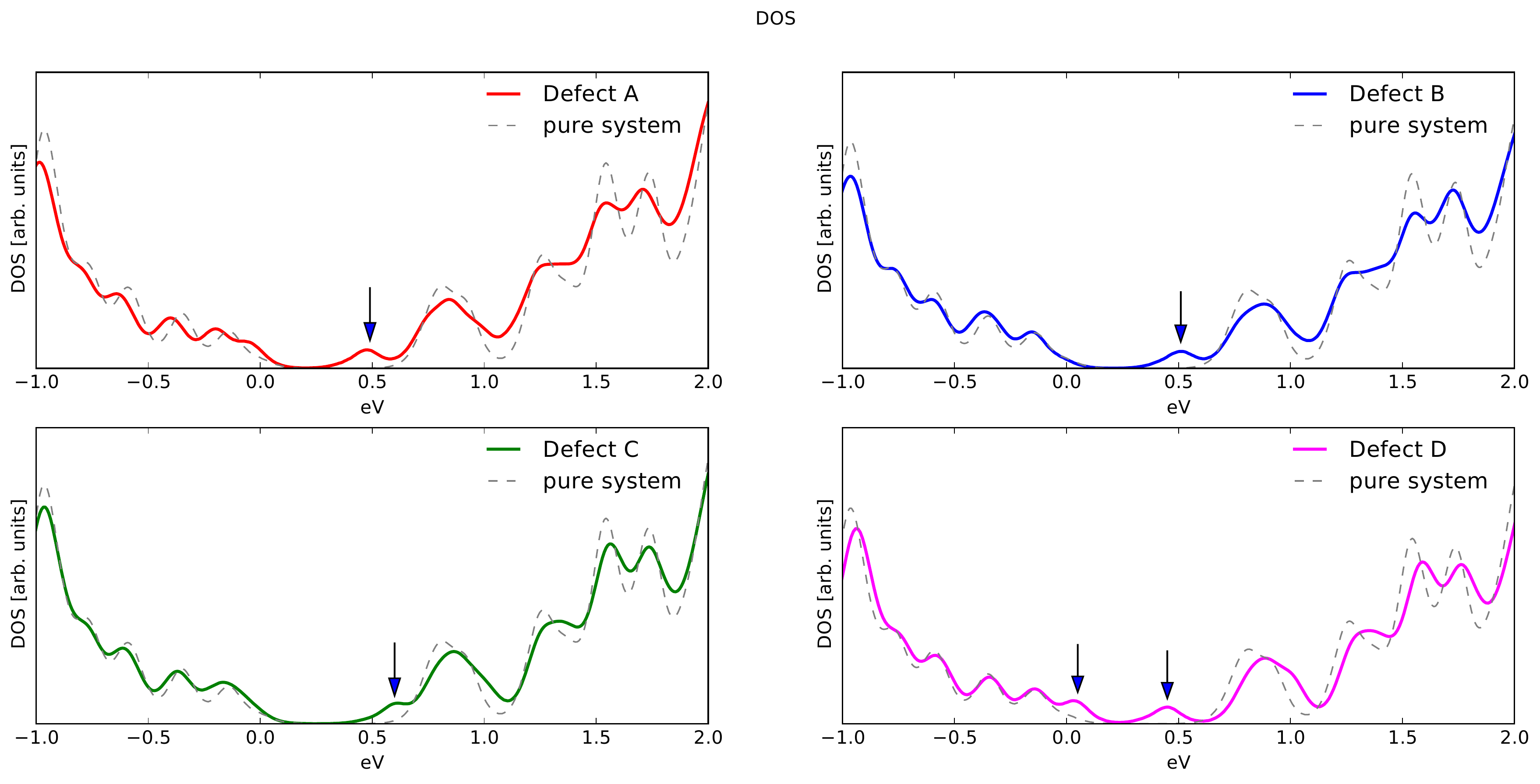}
 \subcaption{\label{fig:defdosip} Density of states in IP approximation}
 \end{minipage}\\
 \begin{minipage}{0.8\textwidth}
  \includegraphics[width=\textwidth]{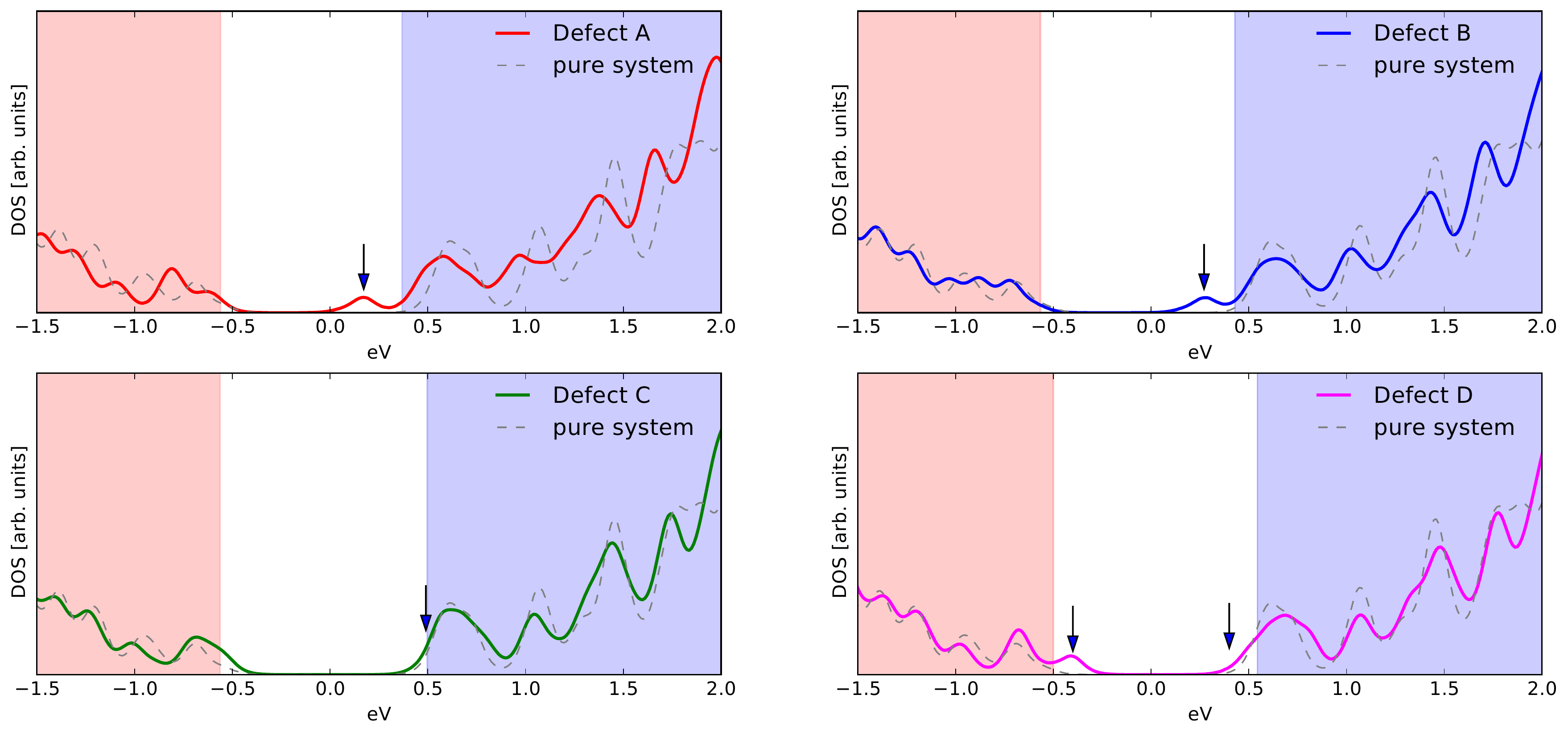}
 \subcaption{\label{fig:defdosgw} Density of states in G$_0$W$_0$ approximation} 
 \end{minipage}
 \caption{[Color online] Top panel: Density of states at the IP level. Bottom panel: Density of states at the G$_0$W$_0$ level, in the presence of the different defect complexes, compared to bulk silicon within the same supercell. The arrows indicate the position of the deep-level defects in the band gap. The red (blue) box indicate the valence (conduction) bands. }
\end{figure}

These observations are strong evidence for the presence of the C-form in irradiated silicon, but more detailed analysis of other properties will be required to definitively assign the observed properties of the dicarbon pair to any single form. 
 Hereafter, we will consider all four forms of C$_i$C$_s$ pair, namely A-, B-, magnetic C-, and D-form, as they are the most interesting configurations. In particular, we will define the proportion of their vibrational and excited states.

 \paragraph{Vibrational properties}
 One of the easiest ways to decipher the precise configuration of defects present in a sample by combined theoretical-experimental investigation of LVMs. 
 Convergence between the values obtained unambiguously indicates a correct structure, while the line-intensity can be used to estimate concentration.
 Therefore, we present our theoretical investigation of LVMs for the four C$_i$C$_s$ forms and compare them with already published results (see Table \ref{tab:frequencies}).
 Our values for the A- and B- forms are in excellent agreement with experimental values and previous calculations.
 For the C-form, our results differ to a larger extent. These differences may be related to the geometries of the C-form studied, which were not the same here and in Docaj's  study.\cite{Docaj2012} Spin polarization was not taken into account by Docaj, as a result, they could be dealing with the D-form or another local minimum.
 The results obtained could serve as a reference for future IR experiments. While the A and B forms present vibrational peaks at no more than $950 \text{ cm}^{-1}$, we propose to justify the existence of C or D forms by the appearance of a peak in the $1100$ to $1200 \text{ cm}^{-1}$ range. Up to now this peak was not observed since, like in the study of Lavrov \textit{et al.} \cite{Lavrov1999}, the spectral range is often only scanned up to $1000 \text{ cm}^{-1}$.
 
The experimental data presented in Fig.\ref{fig:kmal} rely on a 546 cm$^{-1}$ band to measure the C$_i$C$_s$ complex concentration. Our vibrational analysis indicates that this band can be attributed to the B-form or the C-form and the (less stable) D-form. The A-form does not have a band at this frequency. Therefore, combining the vibrational simulations with the KMAL simulations, it appears that the C-form was detected in the experimental studies we examined \cite{Sgourou2013, Londos2013}. 
 
 \begin{table*}[th]
\centering
 \begin{tabular}{cccc ccc ccc ccc}
 \hline
 \multicolumn{4}{c}{This study} & \multicolumn{3}{c}{Docaj \cite{Docaj2012}} & Leary \cite{Leary1997} & \multicolumn{2}{c}{Capaz \cite{Capaz1998}} & \multicolumn{2}{c}{Lavrov \cite{Lavrov1999}} & Lavrov \cite{Lavrov2000} \\ 
 \multicolumn{4}{c}{theory} & \multicolumn{3}{c}{theory } &  theory  & \multicolumn{2}{c}{theory} & \multicolumn{2}{c}{experiment} & experiment \\ 
  \hline
 \textbf{A}   &  \textbf{B}  & \textbf{C-SP} & \textbf{D}    & A   & B  & C-NSP & B   &  A  & B   & A   & B   & not identified\\ 
 \textbf{933} & \textbf{819} & \textbf{1135} & \textbf{1182} & 917 & 805 & 1181 & 838 & 890 & 841 & 953 & 842 & 749 \\ 
 \textbf{861} & \textbf{702} & \textbf{801}  & \textbf{744}  & 912 & 704 & 810  & 715 & 874 & 716 & 873 & 730 & 527 \\ 
 \textbf{699} & \textbf{608} & \textbf{733}  & \textbf{732}  & 710 & 663 & 806  & 649 & 722 & 643 & 722 & 641 & \\ 
 \textbf{572} & \textbf{548} & \textbf{549}  & \textbf{525}  & 598 & 567 & 580  & 582 & 567 & 567 & 597 & 580 & \\ 
 \textbf{566} & \textbf{525} &               &               & 591 & 563 &      & 552 & 557 & 514 & 594 & 543 & \\ 
              & \textbf{521} &               &               &     & 549 &      & 543 &     & 503 &     & 540 & \\ 
\hline
 \end{tabular}
 \caption{Frequencies in cm$^{-1}$ of the four A-, B-, C-, and D-forms of C$_i$C$_s$ pair. Frequencies in bold are calculations from the present work; they are compared to previously-published theoretical and experimental values. In Lavrov \textit{et al.} \cite{Lavrov2000}, the authors did not match frequencies to the form of the C$_i$C$_s$ complex. }
 \label{tab:frequencies}
\end{table*}
 
\paragraph{Electronic and optical properties} 
The dicarbon pairs are associated with a light-emitting defect, the G center. The electronic structure and light spectrum associated with each of the forms described could provide information on which one is linked to the G center. 
We start our discussion of the electronic properties of the four complexes at the Kohn-Sham level. In Fig.~\ref{fig:pbebands}, a schematic representation of the electronic band structure of the four defects is shown. Although we used a large supercell (more than 200 atoms), due to resonant defect states, the presence of the defects slightly affects the bulk gap, shifting it by about 0.13~eV in the worst case - complex A. All four complexes produced deep energy levels in the band gap. In the A, B cases, a single level is present within the gap, located at 0.36~eV, 0.40~eV, respectively, from the top valence band. The C- and D-forms produced more complex structures. In the case of the C-form, the defect was spin-polarized and the level presented in the gap splits into two levels, one for each spin polarization. Finally, the D-form is also associated with two levels in the gap: one fully occupied and one empty. 
Next, we used quasi-particle correction to obtain the band structure schematically represented in Fig.~\ref{fig:gwbands}.

The GW corrections broaden the gap from 0.76~eV to 1.1~eV in pure Si. These corrections increase the energy of the unoccupied defect levels while decreasing that of the occupied levels. In the A and B cases, these levels remain  within the gap whereas in the C and D cases the defect levels almost merge with the lowest conduction and top-most valence lines (see Figs.~\ref{fig:pbebands} and ~\ref{fig:defdosgw}). This difference has important implications for the optical properties, as we will see below. 
The corresponding density of states (DOS), interpolated on $2 \times 2 \times 2$ shifted k-points, is reported in Fig.~\ref{fig:defdosgw}. The DOS of the four C$_i$C$_s$ structures was compared with that of the pure system, aligning the top valence band position. The four complexes share a similar DOS state to the pure system, but the peaks belonging to the bulk silicon are smoother because the defects break symmetries in the supercell. In Fig.~\ref{fig:defdosgw}, the arrows indicate the positions of the defect levels shown in the diagram in Fig.~\ref{fig:gwbands}. Now that we have analyzed the electronic structure of the four defects, we can move on to their optical properties.

\begin{figure}[ht]
 \centering
 \begin{minipage}{8cm}
  \includegraphics[width=\textwidth]{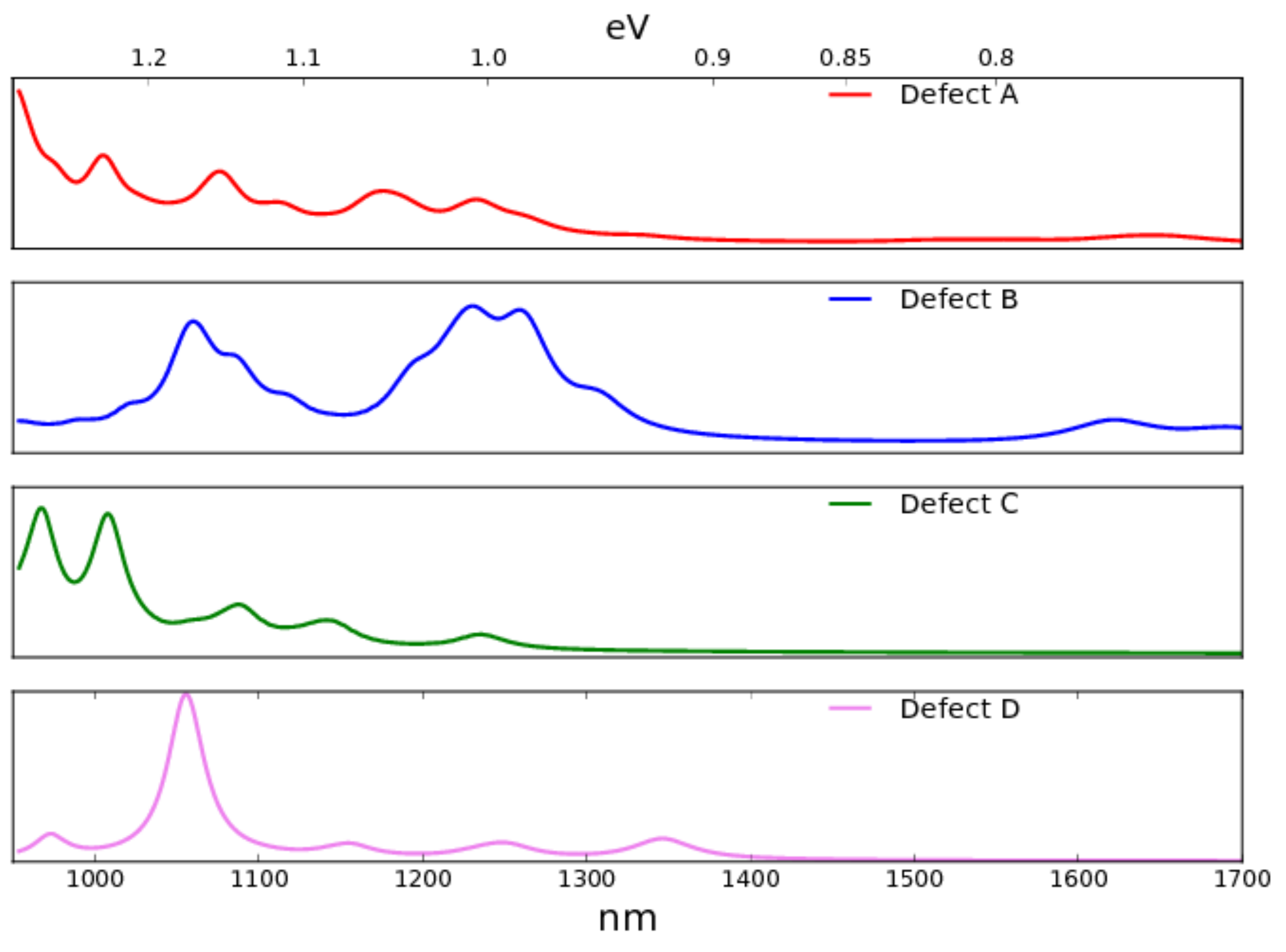}
 \subcaption{\label{fig:optics_ipgw} G$_0$W$_0$ + IP }
 \end{minipage}\\
 \begin{minipage}{8cm}
  \includegraphics[width=\textwidth]{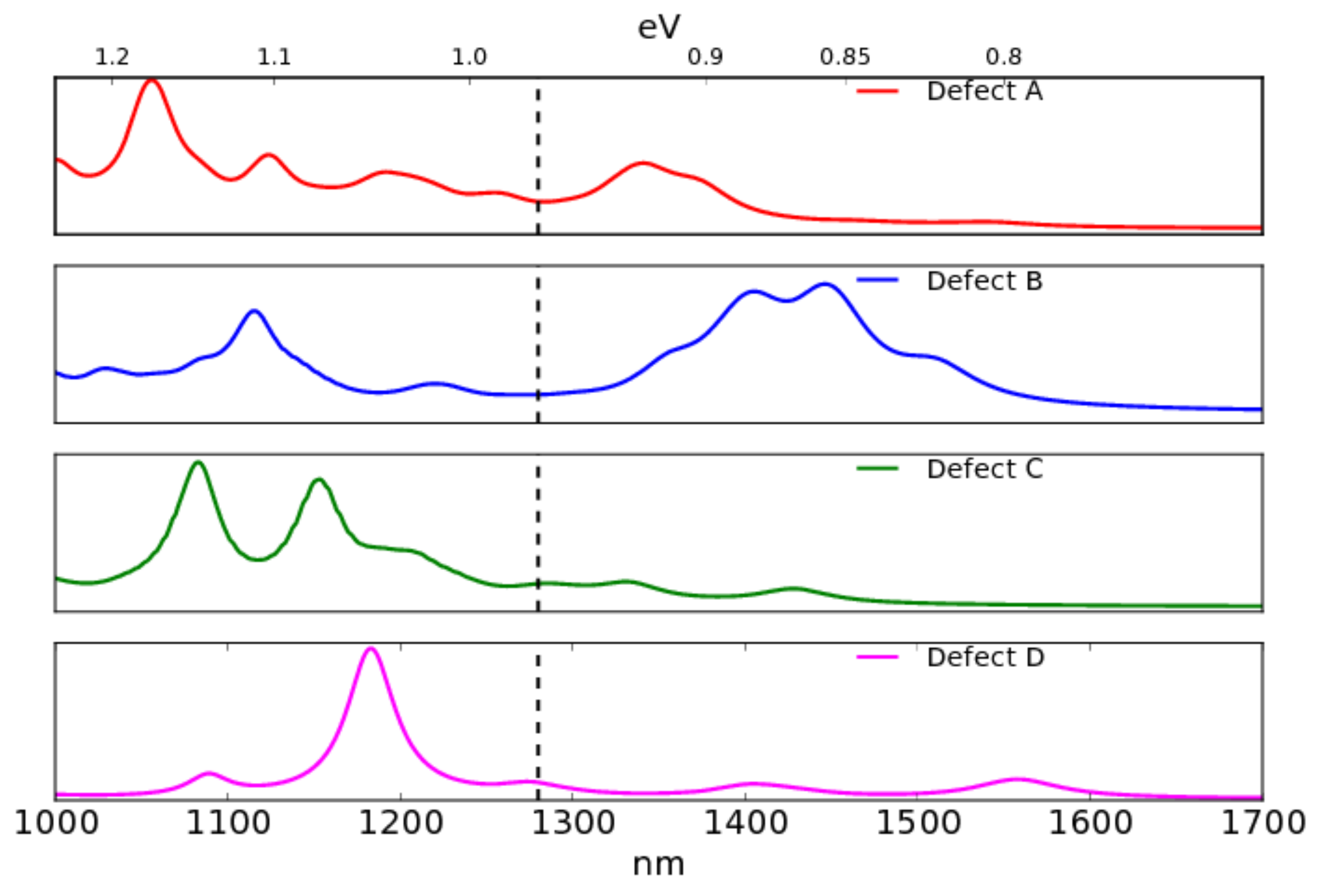}
 \subcaption{\label{fig:optics_bse} G$_0$W$_0$ + BSE} 
 \end{minipage}
 \caption{[Color online] Top panel: Optical absorption in the presence of the different defect complexes in independent particle approximation, starting from the G$_0$W$_0$ band structure. Bottom panel: Optical absorption with the different defect complexes, applying the G$_0$W$_0$ approximation plus the Bethe-Salpeter equation. The vertical lines indicate the wavelength of experimental light emission\cite{Berhanuddin2012a, Berhanuddin2012b} }
\end{figure}

The optical response was evaluated within independent particle approximation (IP) starting from the G$_0$W$_0$ band structure (Fig.~\ref{fig:optics_ipgw}) and by means of the Bethe-Salpeter equation (BSE), \textit{i.e.} including both local-field effects and electron-hole interaction (Fig.~\ref{fig:optics_bse}).
In the IP approximation the optical response can simply be constructed from transitions between occupied and unoccupied levels depicted in Fig.~\ref{fig:gwbands} mediated by the optical matrix elements.
In Fig.\ref{fig:optics_ipgw} we report only the lowest part of the spectra, that is the one more affected by the presence of defects.

Next, we included correlation effect using the Bethe-Salpeter Equation that mixes the single particle transitions, redistributes the spectral weight and renormalizes the transition energies. 
In the B-, C-, and D-forms of C$_i$C$_s$ complex, these effects simply shift the spectra towards lower frequencies (see the difference between Fig.~\ref{fig:optics_ipgw} and Fig.~\ref{fig:optics_bse}). In contrast, in the case of the A-form, the spectral weight is significantly redistributed, the peak at 1300~nm disappears and the peak at 1350~nm becomes more intense. 
%
%
Fig.~\ref{fig:optics_bse} shows the final optical absorption spectra for the four defect forms. Note that the spectra for the A, B and C complexes are of similar intensity, whereas the spectrum for the D-form is four times more intense.
This can be explained by the fact that both occupied and unoccupied levels of the D-defect are present in the band-gap, and therefore transition between them have a larger weight in the spectra due to the small energy difference.\\
Comparing optical absorption results with luminescence measurements is not easy for several reasons. Firstly, the intensity of the two spectra is unequal due to the different electron distributions. In addition, the luminescence spectrum is usually red-shifted relative to the absorption spectrum due to structural relations (Stoke shift), and this shift has similar energy to that of local vibrations.
For these two reasons, it is challenging to identify particular defects from their optical spectra. 
From the optical response presented here, the defect complexes that could be excluded were the A and B complexes, because they cause strong absorption peaks at wavelengths below emission line. Thus, once excited, a defect like the B (or the A),  will emit in a range above 1450 nm (below 0.8 eV).
As a result, the C- and D-forms appear to be the most likely to emit light. Moreover, the energy shift between their optical absorption peak and the emission peak is compatible with local vibration energies (see Table~\ref{tab:frequencies}). 
Finally, the C and D cases also produced small peaks at a lower energy than the emission peak, and these could come into play in the luminescence process. It is therefore challenging to identify the defect complex from our absorption spectra.

\section{Conclusions}
\label{conclusions}

In this study we performed a detailed investigation into the properties of the forms of the C$_i$C$_s$ complex from first principles calculations.
 Our aim was to better understand the properties of the possible forms of  C$_i$C$_s$ complexes and why the C-form, which is the most stable according to our calculations, has never yet been experimentally observed. In addition, we tried to theoretically characterize the optical properties of the four carbon-carbon pair forms, and attempted to assign one of them to the light-emitting G-center.
 
 Our results indicate that, among all four C$_i$C$_s$ forms, the C-form is the most stable, with binding energy 0.4 eV higher than that of the B-form. Moreover, KMAL isochronal annealing simulations demonstrated that the dissociation temperature for the 546~cm$^{-1}$ band, which is used to determine the stability of C$_i$C$_s$ complex, corresponds to the binding energy of the C-form. Meanwhile, the binding energy of the B-form results in a dissociation at a temperature about 100 \textcelsius ~lower, which strongly contrasts with experimental data and therefore further supports the existence of the C-form. 
 We next computed a set of localized vibrational modes for each of the four configurations. The set corresponding to the C-form contains four bands, three of which are within the same range as the A- and B-forms, and even overlap with the latter. The highest band, at 1135 cm$^{-1}$, is above the registration range in most LVM experiments, and it could therefore have been missed during measurements.
 The electronic and optical properties of the C$_i$C$_s$ complexes indicated that correlation effects must be included to describe the optical properties of C$_i$C$_s$ defect complexes. In fact, the various approximations in standard semi-local functionals fail to describe the appropriate position level for localized and resonant defect states with respect to the bulk levels.
 Our results thus provide an accurate quasi-particle band structure for the four complexes and their optical absorption using the GW+BSE approximation.
 However, as a consequence, the form responsible for the light-emitting G-center becomes ambiguous. The C-form seems to be the most stable one, but both the B- and C-forms are compatible with the vibrational measurements. Nevertheless, the optical response tends to exclude the A and B complexes.
But defects A and B could contribute to the luminescence process through non-radiative decay. This possibility was not considered in this work. 
To summarize, all four forms can exist and probably coexist in carbon-rich irradiated silicon. The relative concentrations of the four complexes probably depends on their binding energies if the sample is at thermal equilibrium, or,  otherwise, on the thermal history of the sample.
It is difficult to identify a single form which would be responsible  for the light emission of the G-center in silicon.
However, the stability, vibrational properties and optics provide strong evidence that the C-complex plays an important role  in heavily carbon-doped silicon.
 To our knowledge, no evidence of the existence of the C-form has been obtained by, the generally effective EPR characterization technique. According to our simulations, the C-form is a magnetic complex, and as a result it would produce a signal in a spectral range far from the signal for neutral species.
 Hence, additional characterization experiments will be required to investigate the C and D configurations. New experiments should take into account the complexes' vibrational properties, as reported in the current study, as well as their magnetization. From a theoretical point of view, some additional calculations of excited states of C$_i$C$_s$ forms could provide new information. 

\section{Acknowledgments} 
The work was funded by ANR as part of the BOLID project (ANR-10-HABI-0001).
Computing time was provided by the national GENCI-IDRIS and GENCI-TGCC supercomputing centers under contracts $n^o$ t2012096655 and $n^o$ t2014096107.
\addcontentsline{toc}{chapter}{Bibliography}
\bibliographystyle{apsrev4-1}

\begin{thebibliography}{33}
\bibitem [{not()}]{noteopt}%
  \BibitemOpen
  \href@noop {} {}\bibinfo {note} {We checked that the optimized structures did
  not significativity change if we reoptimize them using plane-waves and
  Troullier-Martins pseudopotentials.}\BibitemShut {Stop}%
\end{thebibliography}

\end{document}